\documentclass[aps,prb,amssymb,twocolumn,showpacs]{revtex4}

\usepackage{graphicx}

\def \be{\begin{equation}}
\def \ee{\end{equation}}
\def \ba{\begin{array}}
\def \ea{\end{array}}

\def \etal{{\it {et al}}}

\def \br{{\bf r}}

\def \bk{{\bf k}}

\def \bx{{\bf x}}
\def \bnull{{\bf 0}}

\newcommand{\bea}{\begin{eqnarray}}

\newcommand{\eea}{\end{eqnarray}}

\newcommand{\beq}{\begin{equation}}

\newcommand{\eeq}{\end{equation}}

\newcommand{\lav}{\langle}

\newcommand{\rav}{\rangle}

\newlength{\textwidthm}

\begin{document}

\title{Light cone dynamics and reverse Kibble-Zurek mechanism in two-dimensional
 superfluids following a quantum quench}

\author{L.~Mathey$^1$ and  A.~Polkovnikov$^2$}

\affiliation{$^1$Joint Quantum Institute, National Institute of Standards and Technology and University of Maryland,
 Gaithersburg, MD 20899\\
$^2$Department of Physics, Boston University, 590 Commonwealth Ave., Boston, MA 02215}

\date{\today}

\begin{abstract}
  We study the dynamics of the relative phase of a bilayer of
  two-dimensional superfluids after the two superfluids have been
  decoupled. We find that on short time scales the relative phase
  shows ``light cone'' like dynamics and creates a metastable
  superfluid state, which can be supercritical.  We also demonstrate
  similar light cone dynamics for the transverse field Ising model. On
  longer time scales the supercritical state relaxes to a disordered
  state due to dynamical vortex unbinding.  This scenario of
  dynamically suppressed vortex proliferation constitutes a {\it
    reverse-Kibble-Zurek effect}.  We study this effect both
  numerically using truncated Wigner approximation and analytically
  within a newly suggested time dependent renormalization group
  approach (RG). In particular, within RG we show that there are two
  possible fixed points for the real time evolution corresponding to
  the superfluid and normal steady states. So depending on the initial
  conditions and the microscopic parameters of the Hamiltonian the
  system undergoes a non-equilibrium phase transition of the
  Kosterlitz-Thouless type. The time scales for the vortex unbinding
  near the critical point are exponentially divergent, similar to the
  equilibrium case.
\end{abstract}


\maketitle

\section{Introduction}
The technological advances of trapping and manipulating ultra-cold
atom systems provide an opportunity to study many-body dynamics with
unprecedented clarity.  The realization of Bose-Einstein condensates
in ultra-cold atom systems \cite{BEC-exp}, the Mott insulator
transition~\cite{greiner}, the BEC-BCS transition \cite{BCS-BEC}, the
Kosterlitz-Thouless transition \cite{BKT, zoran, clade}, demonstrated
that this technology can be used as a quantum simulator of many-body
phases. Here, the static state of a system in equilibrium is created
and studied. Various dynamical aspects of ultra-cold atom systems have
also been probed, such as dipole oscillations\cite{dipole}, vortex
excitations~\cite{vortex}, and soliton dynamics~\cite{soliton},
absence of equilibration in one-dimensional bosonic
systems~\cite{newtoncradle}, spontaneous formation of vortices in
spinor condensates~\cite{sadler} and many others (see
Ref.~[\onlinecite{bloch_review}] for a recent review).  In
Ref.~[\onlinecite{stirring}], vortices excitations were created via
laser stirring. In these experiments, the dynamics of only a few
degrees of freedom were studied, such as the center of mass motion, or
the dynamical evolution of a vortex. These experimental developments
stimulated a considerable theoretical interest in understanding
non-equilibrium quantum dynamics including analysis of dynamics
following sudden quenches~\cite{quench}, studying connections between
dynamics and thermodynamics~\cite{therm}, dynamics through quantum
critical points~\cite{adiabatic}.

The focus of this paper is a detailed analysis of the full many body
dynamics following the quench in a two-dimensional quantum rotor
model. Physically we imagine the situation where two initially
strongly coupled superfluids are suddenly separated and we are
interested in the evolution of the relative phase between the two
superfluids. In particular, we will be interested in the question of
how the system relaxes to the equilibrium state.  We note that
experiments in a similar setup involving separation of two 1D
superfluids were reported in Ref.~[\onlinecite{joerg1}] and the
corresponding theoretical analysis was done in
Refs.~[\onlinecite{burkov, rafi, mazets}]. Unlike the 2D case,
phonon fluctuations in 1D result in the exponential decay of the
correlation functions and nonlinear effects in the form of phase slips
do not bring qualitative changes to the behavior of the correlation
functions at least at low initial temperatures~\cite{rafi}.

In equilibrium for the uncoupled layers, there are two possible
phases. At low temperatures atoms in each layer (which we regard as
identical) form a (quasi-)superfluid phase while at high temperatures
they form a normal Bose gas. These phases can be distinguished by the
long-range behavior of the single particle correlation function
$G(\bx) = \langle b^\dagger(0) b(\bx)\rangle\approx \rho \langle
\exp[i(\phi(\bx)-\phi(0))]\rangle$, where $b(\bx)$ is the single
particle operator, $\rho$ is the atom density, and $\phi$ is the
phase. We note that a rotor representation of bosons $b(\bx)\sim
\sqrt{\rho(\bx)}\exp[i\phi(\bx)]$ is possible when the healing length
characterizing the characteristic length scale of density fluctuations
is short compared to other length scales in the problem. Under the
same conditions the density fluctuations are negligible if we are
interested in long distance physics. In the superfluid phase this
function shows algebraic scaling, $G(\bx) \sim |\bx|^{-\tau/4}$, where
the scaling exponent $\tau$ is proportional to the temperature
$\tau\approx T/T_c$, with $T_c$ being the Kosterlitz-Thouless
temperature. At the transition point we have $G(\bx)\sim|\bx|^{-1/4}$.
Above the transition, the correlation function shows exponential
scaling $G(\bx) \sim \exp(-|\bx|/\xi)$, with some correlation length
$\xi$ which diverges near the transition temperature.  The algebraic
scaling of the superfluid phase is due the thermally excited phonon
(Bogoliubov's) modes.  In two dimensions these fluctuations generate
quasi-long range order, rather than true long range order.  The
transition to the exponential regime is due to vortex excitations.
Above the transition, vortex-antivortex pairs are deconfined so that
vortices and anti-vortices become unbound. These excitations generate
a much more disordered phase field, which leads to exponential scaling
of the correlation function.

If we couple the two superfluids with a hopping term in the
temperature regime of the critical temperature, the system forms a
phase-locked state, see Refs.~[\onlinecite{mathey07c, cazalilla}].
Here the correlation function of the relative phase scales as $G(\bx)
\sim \exp(-|\bx|/\xi_{i}) + C$, where $C\neq 0$, and $\xi_i$ is the
correlation length of the phase-locked state.  In this state the
relative phase is well-aligned over long distances; its fluctuations
are strongly suppressed.  We then turn off the hopping and study the
evolution of the system.

In this paper we show that the relaxational dynamics occurs in two
stages. The first fast stage, which we will term light cone
relaxation, establishes a metastable quasi-equilibrium state of
phonons (or Bogoliubov excitations) characterized by effective
non-equilibrium temperature (in principle this metastable state can be
completely non-thermal). During this stage the correlations between
two arbitrary points in space $\bx_1$ and $\bx_2$, $G(\bx_1,\bx_2,t)$,
where $t$ is the time after the quench, initially decay in time, independent of
their spatial separation $x=|\bx_1-\bx_2|$ because these points are
not causally connected: $G(\bx_1,\bx_2,t)\sim 1/t^\alpha$, where
$\alpha$ is a power-law exponent related to the parameters of the
system. At a later time $t^\star$, when the condition $2vt^\star=x$ is
fulfilled, these correlations (approximately) freeze in time so that
$G(\bx_1,\bx_2,t)\sim 1/x^\alpha$. The exponent $\alpha$ thus
defines the non-equilibrium phonon temperature in the system. Because this
first stage of dynamics involves only phonons, the exponent $\alpha$
can exceed the maximally allowed equilibrium value of one fourth, leading
to a non-equilibrium super-critical metastable state, which can be
thought of as a supercritical superfluid. It is analogous to an
overheated classical liquid, for which a liquid state can be sustained
above the critical temperature if the creation of defects is
avoided. We find that the power-law can be substantially above the
critical scaling, and furthermore, that this metastable can be very
long-lived.

At longer time scales, vortex-antivortex pairs emerge and proliferate
leading to the true equilibrium state. This process occurs at much
longer time scales. We describe this thermalization process both
numerically, using truncated Wigner approximation (TWA) and
analytically. In particular, we show that thermalization (here
corresponding to the process of vortex-antivortex proliferation) can be
understood by extending renormalization group ideas to real time
dynamics. By doing partial averaging over fast oscillating high energy
degrees of freedom, we can rewrite the equations of motion of slower
degrees of freedom through renormalized coupling constants. As in
the case of equilibrium systems we observe two possible scenarios
corresponding to vortex-antivortex pairs being irrelevant (superfluid
phase) or relevant (normal phase). Thus we are able to see how the
system relaxes to one of the phases in real time. Divergent time (and
length) scales in equilibrium systems translate into divergent
relaxation times required to reach thermalization in the
non-equilibrium case.

\begin{figure}
\includegraphics[width=7.0cm]{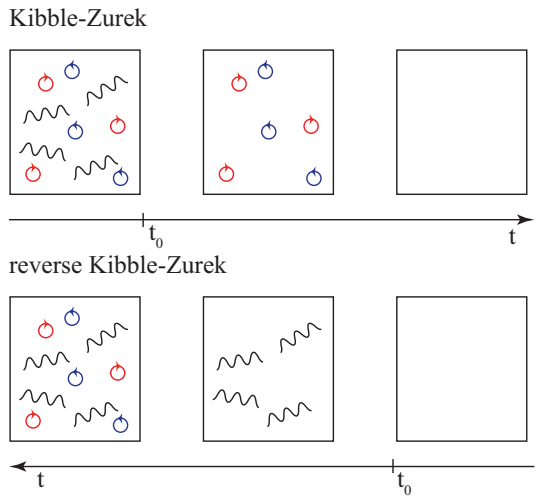}
\caption{Illustration of the Kibble-Zurek (KZ) mechanism, which describes  ramping across a phase transition from the disordered phase, and  its counterpart, the reverse-Kibble-Zurek (rKZ) effect.  The latter describes ramping across a transition from the ordered side.  Its defining feature is the dynamical suppression of vortex unbinding,  which happens on a much longer time scale than the appearance of phononic  excitations. We propose  to study the rKZ in a bilayer of 2D superfluids of ultra-cold atoms,  by decoupling the superfluids and measuring the dynamics
 of the relative phase.}\label{KZsketch}
\end{figure}

Physically this decay of the metastable superfluid state to the new
equilibrium is very reminiscent of the Kibble-Zurek (KZ) effect.  The
latter describes a ramp across a phase transition, starting on the
disordered side. If the ordered state supports topological excitations, 
like vortices, then one expects very slow relaxation of the resulting
state to the equilibrium due to vortex-antivortex recombination.  This
scenario is illustrated in Fig. \ref{KZsketch} a): In the disordered
phase we have excitations such as phonons, as well as topological
defects.  When we apply a fast ramp across the phase transition, the
phonon excitations thermalize on very short time scales, while
topological defects can exist on much longer time scales. The
mechanism of relaxation in our case is exactly complimentary and can
be termed as reverse Kibble-Zurek effect. Here, the ramp across a
phase transition starts from the ordered side, as illustrated in
Fig. \ref{KZsketch} b). In the ordered phase both phonon excitations
and vortices are suppressed. When the system is ramped across the
transition, phonons are generated on a fast time scale. However,
vortices are generated at much longer time scales leading to the
long-lived supercritical superfluid state. We point out that in
thermally isolated systems (like cold atom systems) it is much easier to
observe reverse KZ effect because the disordered phase usually corresponds
to a higher temperature. In isolated systems it is relatively easy to
increase temperature by quenching some parameter, while decreasing
temperature requires much more effort and can be done only in open
systems.

This paper is organized as follows: In Sect. \ref{TWA} we introduce
the numerical method that we use and find that at short time scales
the system shows light cone dynamics. In Sect. \ref{lindyn} we
consider the linearized dynamics of the bilayer system. Within this
approximation both light cone dynamics and the emerging superfluid
state can be understood. In Sect. \ref{tIsing} we study the light cone
dynamics of a solvable model, the transverse Ising model. In
Sect. \ref{vunbind} we study dynamical vortex unbinding both with
truncated Wigner approximation, and with a renormalization group
approach. We note that a short version of this paper with some of the
results was published earlier~\cite{LM_AP_01}. Here we expand the
earlier treatment, present additional results and derivations, and
formulate the real time renormalization group approach which explains the
numerical results.

\begin{figure}
\includegraphics[width=7.0cm]{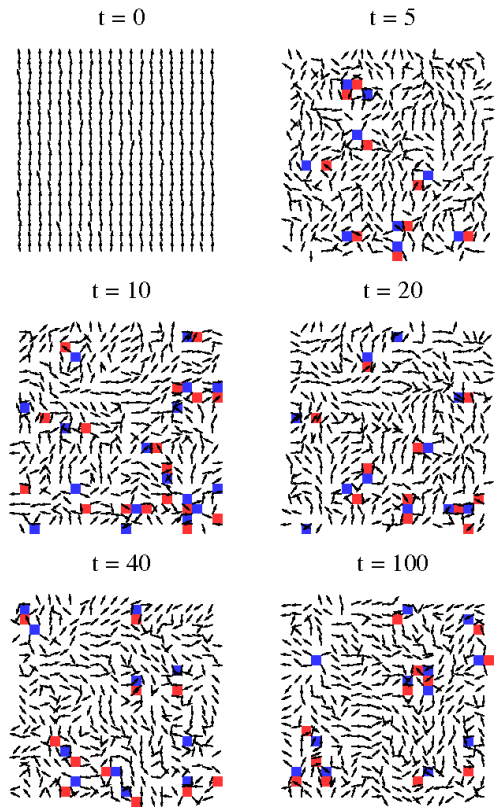}
\caption{\label{run} We simulate the dynamics of the relative phase of
  two 2D superfluids by solving the equations of motion and by
  averaging over the Wigner distribution of the initial state. A
  single run is shown here, for $V=100$, $\kappa=10$ and $T=2$, at the
  times $t = 0, 5, 10, 20, 40, 100$.  Vortices are marked red,
  anti-vortices blue.  }
\end{figure}

\section{Microscopic Model and the Truncated Wigner Approximation}\label{TWA}
In this section we present the model that we use in our numerical
approach.  We consider two two-dimensional (quasi-)condensates that
are aligned in parallel to each other, that are coupled by a hopping
term which is then turned off.  This can be achieved by increasing the
potential between the two condensates. The coarse-grained Hamiltonian
describing the relative phase $\phi_i$ of the two superfluids
corresponds to an XY model, to which we add a hopping term to describe
the phase-locking in the initial state:
\bea
H & = &  \Omega_0\Big(-\sum_{<ij>}\frac{\kappa}{\pi}\cos(\phi_i-\phi_j)
 + \frac{\pi}{2 \kappa} \sum_{i} n_{i}^2\nonumber\\
& & - V(t) \sum_{i} \cos(\sqrt{2}\phi_i)\Big),
\label{hamiltonian}
\eea
where $\Omega_0$ is an overall (Josephson) energy scale, $\kappa$
describes the ratio of kinetic and potential energies.  We can
formally replace these parameters by $\Omega_0\kappa/\pi = 2 J n$,
$\pi \Omega_0/\kappa = U$ (so that $\Omega_0=\sqrt{2JnU}$,
$\kappa=\pi\sqrt{2Jn/U}$) and $V(t) = 2 J_\perp(t) n/\Omega_0$, which
gives a representation of two coupled Bose-Hubbard systems in the
quantum rotor limit~\cite{sachdev}. In this limit the Bose operators
 are replaced by  the phase-density representation and
the fluctuations of density are assumed to be small.  In the
Bose-Hubbard model $J$ is the in-plane hopping amplitude, $U$ is the
on-site interaction energy, $n$ is the filling number, i.e. the number
of particles per site, and $J_\perp$ is the inter-layer hopping
amplitude $J_\perp$. This representation gives at best a qualitative
idea of how the model parameters relate to the parameters in
experiment, but gives a more intuitive picture. We note that one can
think about the continuum limit as discrete, where the lattice
constant is approximately given by the zero-temperature healing length
in the system, i.e. the length over which density fluctuations are
suppressed.

We emphasize that despite the BKT transition being classical in
origin, i.e. driven by thermal fluctuations, the mechanism of vortex
or phonon creation in the process we consider comes from quantum
fluctuations. Indeed when the superfluids are strongly coupled
together the density (which plays the role of momentum conjugate of 
the phase) strongly fluctuates because of the zero point motion. The
heating mechanism of this system can be thought of as enhancement of
this zero point motion following the quench.

It is convenient to introduce the rescaled quantities $\tilde{t} =
\Omega_0 t/ \hbar$, $\tilde{\phi} = \sqrt{\frac{\kappa}{\pi}} \phi$,
and $\tilde{n} = \sqrt{\frac{\pi}{\kappa}} n$.  In terms of these, the
classical equations of motion (EOMs) are
\bea
\frac{d\tilde{\phi}_i}{d\tilde{t}} & = & - \tilde{n}_i\\
\frac{d\tilde{n}_i}{d\tilde{t}} & = & -
\frac{\sqrt{2}}{\beta}\sum_{j_i}\sin\Big(\frac{\beta(\tilde{\phi}_{j_i}-
  \tilde{\phi}_i)}{\sqrt{2}}\Big) + V(t) \beta \sin \beta
\tilde{\phi}_i,
\label{eom}
\eea
where we defined $\beta = \sqrt{2\pi/\kappa}$. The indices $j_i$
describe the four neighboring sites of site $i$.

We model the relative phase using a numerical implementation of the
truncated Wigner approximation (TWA) (see Ref.~[\onlinecite{blakie}]
for a review): The expectation of any quantity at some time $t>0$
can be determined by sampling over a Wigner distribution at time
$t=0$, and solving the classical equations of motion from $0$ to
$t$. This approximation is guaranteed to be accurate at short
times~\cite{ap_twa, ap_long}. This approximation is also exact for any
quadratic theory so we expect it to be accurate in the first
(light-cone) stage of dynamics primarily driven by phonon
excitations. In our case we expect that TWA is also valid at longer
times because when vortex-antivortex pairs start to emerge the system
already reached metastable state corresponding to finite effective
temperature. At this point quantum fluctuations become suppressed by
much stronger thermal fluctuations driving the slow vortex dynamics.

We solve these EOMs for initial conditions that are distributed
according to the Wigner distribution at $t=0$.  We can calculate this
distribution under the assumption that $J_\perp$ is larger than the
other energy scales at $t=0$.  In this limit the phase fluctuations
are small and can be described within the Bogoliubov approximation,
where the system reduces to a sum of oscillators.  The Fourier modes
$\tilde{\phi}_q$ and $\tilde{n}_q$ at $t=0$ are distributed according
to (see Ref.~[\onlinecite{ap_long}])
\bea
W & \sim & \exp\Big(-\frac{\tilde{\phi}_q^2}{2 \sigma_q^2 r_q}
 -\frac{2 \sigma_q^2\tilde{n}_q^2}{r_q}\Big),
 \label{wig}
\eea
where $\sigma = 1/\sqrt{2\omega_q}$, $r_q = \coth(\omega_q/2T_0)$, and
$\omega_q = \sqrt{4 \sin(q_x/2)^2 + 4 \sin(q_y/2)^2 + V\beta^2}$ with
$T_0$ being the initial temperature. Note that formally $\omega_q$
diverges at $V\to\infty$. This divergence is unphysical, being an
artifact of using Hamiltonian~(\ref{hamiltonian}) in the number phase
representation. In reality when $J_\perp$ becomes very large the
transverse Josephson frequency saturates at $\omega\approx
2J_\perp$. This happens at $V\sim n$ or equivalently $J_\perp\sim
Un$. So for very strong initial coupling one can still use
distribution~(\ref{wig}) with $V\to n$.

To visualize our simulations we show an example for a single run of
the system on a 20-by-20 lattice in Fig.~\ref{run}.  The direction of
the arrows on each lattice point describe the phase $\phi_i$. We show
'snapshots' at various times.  The plaquettes around which there is a
phase winding of $\pm 2\pi$ are marked as vortices and
anti-vortices. We see that at $t=0$, the phases are well aligned due
to the coupling between the layers, with some small quantum
fluctuations described by the Wigner function.  The coupling is then
turned off, vortices and anti-vortices are created pair-wise, and
unbind on a long time scale, as we will discuss further on.  To
extract expectation values of our observables from our simulations, we
have to average them over many realizations of initial fluctuations.

\begin{figure}
\includegraphics[width=7.2cm]{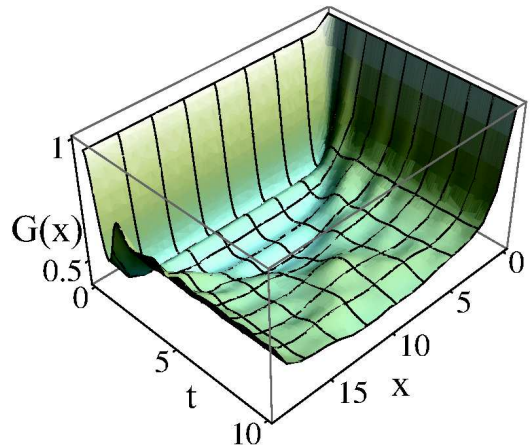}
\caption{\label{GF}
Plot of short-time behavior of
the correlation function as a function of time and space, at temperature
 $T=3$, for $\kappa=10$ and $V=20$. The dynamics separates into instantaneous,
 damped oscillatory behavior, and a 'light cone' like pulse.
}
\end{figure}

We use this method to extract the equal time correlation function:
\beq
G(x,t)=\langle \exp[i\sqrt{2}\phi_j(t)-i\sqrt{2}\phi_{j+x}(t)]\rangle,
\eeq
where $x$ is an integer separation between the points and $t$ is the
time after decoupling (see Fig.~\ref{GF}). Because we are using
periodic boundary conditions $G(x,t)$ depends only on the separation
between the points $x$ and does not depend on $j$. Note that this
correlation function (or rather $\int_0^x dx' G(x',t)$) can be
directly measured in interference experiments~\cite{interference,
  zoran, joerg1}. We indeed see very clear emergence of the light cone
thermalization: At separations larger than $2vt$, where $v$ is
characteristic phonon velocity, $G(x,t)$ is almost $x$ independent -
it uniformly decreases in time. Once $2vt>x$ the correlations freeze
in time and depend only on $x$.  The quantities in the system have
 been rescaled such that the phonon velocity is set to 1.

We find that the state that emerges within the light cone shows
algebraic scaling, and therefore can be referred to as a superfluid.
%
%

\section{Linearized Dynamics}\label{lindyn}
In this section we study the linearized dynamics of the system. Within
this description, both the light cone dynamics and the metastable SF
state can be understood.  The quadratic Hamiltonian describing the
relative phase of two coupled superfluids reads:
\bea
H_{0} & = & \int d^2r \Big(-\frac{v}{2 r_0}(\nabla\phi)^2 + \frac{g_\perp}{2}\phi^2
 + \frac{v r_0}{2}n^2\Big).\label{Hamlincont}
\eea
$v$ is the phonon velocity of the SF, approximately given by
$v=\sqrt{g n/m}$.  $r_0$ is the short-range cut-off of the system, of
the order of the healing length.  $J=v/r_0$ is the KT energy.  The
term $g_\perp\phi^2/2$ is created by the hopping term of the
bilayer. It is approximately given by $g_\perp = 4 J_\perp n$.  We
note that when this Hamiltonian is put on a lattice with a lattice
constant $r_l$  we obtain
\be
H_{1}  =  \Omega_0 \sum_i\Big(-\frac{r_l}{r_0}
\sum_{j_i}\frac{(\phi_i-\phi_{j_i})^2}{2}
+ \frac{g_\perp r_l^2}{2 \Omega_0} \phi_i^2 + \frac{r_0}{2 r_l} n_i^2\Big).
\ee
This expression can be also obtained directly linearizing the original
Hamiltonian~(\ref{hamiltonian}). The index $j_i$ here describes the
four neighboring sites of the site $i$, and $n_i$ is the filling
fraction, related to the density via $n_i = n r_l^2$.  $\Omega_0$ is
related to the phonon velocity as $\Omega_0 = v/r_l$.  We therefore
find that the squeezing parameter $\kappa/\pi$ is given by $\kappa/\pi
= r_l/r_0$, i.e. it is the ratio of the discretization length scale
$r_l$ and the short-range cut-off $r_0$ of the system.
 $g_\perp$ is related to $V(t)$ by $g_\perp r_l^2/2 = \Omega_0 V(t)$.

We now consider the time evolution of $\phi$ and $n$ under
(\ref{Hamlincont}).  It is convenient to go to  momentum representation
where different modes decouple from each other. Assuming also that we
are interested in momenta smaller than $1/r_l$, where the lattice
effects are not important we obtain the following equations of
motion:
\bea
\frac{d}{d t} n_k & = & - \Omega_0 \Big(\frac{r_l \epsilon_k^2}{r_0}
 + 2 V\Big)\phi_{-k}\\
\frac{d}{d t} \phi_k & = & \Omega_0 \frac{r_0}{r_l} n_{-k}.
\eea
where
 $\epsilon_k^2=4\sin^2 k_x/2+4\sin^2 k_y/2$,
 and $k$ is dimensionless, $k = -\pi ...\pi$.
 We rescale the time variable as $\tilde{t} = \Omega_0 t/\hbar$.
 The initial dispersion is then given by
\bea
\omega_{k, 0}^2 & = & \epsilon_k^2  + 2 V r_0/r_l.
\eea
 The dispersion $\omega_k$ after the quench is simply $\omega_k^2 = \epsilon_k^2$,
 in these units.
 We solve these equations
 and
 calculate the equal-time correlation function at time $t$ after the
 quench. We use
\bea
G(\bx, t) & = & \langle\exp(i\phi(\bnull, t))\exp(-i\phi(\bx, t))\rangle\\
 & = & \exp(-\langle\delta\phi^2\rangle/2),
 \label{gxt}
\eea
where $\delta \phi= \phi(\bnull, t)-\phi(\bx, t)$.  The averaging is
now trivially done using the Wigner distribution~(\ref{wig}). If we
put the system back to the lattice we then find
\bea
&&\langle\delta\phi^2\rangle
 = \sum_\bk (2 - 2\cos \bk \bx)\nonumber\\
&&~~~\times \Big(
 \frac{r_{k,0}}{2\omega_{k,0}} \cos^2(\omega_k {t})
 + \frac{r_{k,0} \omega_{k,0}}{2\omega_k^2}\sin^2(\omega_k {t})\Big).
 \label{phiav}
\eea
 The quantities   $r_{k,0}$ and $\omega_{k,0}$  are defined as before.
\begin{figure}
\includegraphics[width=7.0cm]{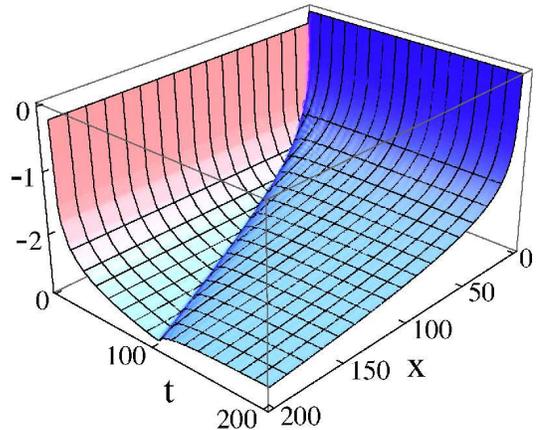}
\caption{\label{phi} $\langle\delta\phi^2\rangle$
 of the linearized system, for $T=1$ and $V\beta^2= 20$, as
 function of the lattice site, and $vt$. }
\end{figure}
\begin{figure}
\includegraphics[width=7.0cm]{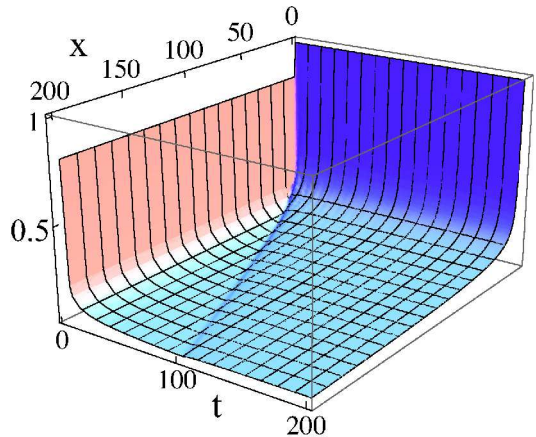}
\caption{\label{GFlin} The correlation function
 of the linearized system, for $T=1$ and $V\beta^2= 20$, as
 function of the lattice site, and $vt$. }
\end{figure}

We now calculate the Green's function in the linearized regime
numerically using Eqs.~(\ref{gxt}) and (\ref{phiav}). We choose the
discretization $r_l = r_0$, the initial temperature $T/\Omega_0 = 1$,
and the initial coupling $V\beta^2= 20$.  In Fig.~\ref{phi} we plot
$\langle\delta\phi^2\rangle$ and in
Fig.~\ref{GFlin} we plot the correlation function.  In both plots the
light-cone dynamics is clearly visible. Because of the translational
invariance the correlation function that emerges in the light-cone
only depends on the relative distance is given by \be G(\bx, t)
\approx C_1 |x|^{-T*/4 T_{KT}}
\label{corr_in}
\ee
 for $x \ll 2 v t$,  where $T^*$ is an effective temperature that is estimated below, and $C_1$ is a numerical prefactor.  Outside of the light cone ($x\gg 2vt$) the function $G(\bf x,t)$ only depends on time $t$ but not on the distance $\bx$:
\be
G(\bx, t)  =  C_2 |t|^{-T*/4 T_{KT}},
\label{corr_out}
\ee
where $T^*$ is the same effective temperature. At the light cone boundary $x\approx 2vt$ the two asymptotics for the correlation function (\ref{corr_in}) and (\ref{corr_out}) approximately coincide. However, we note that the prefactor $C_2$ is in general different from $C_1 v^{-T*/4 T_{KT}}$ as it is evident from the existence of a wavefront that is visible in Figs.~\ref{phi} and \ref{GFlin}.

\begin{figure}
\includegraphics[width=8.8cm]{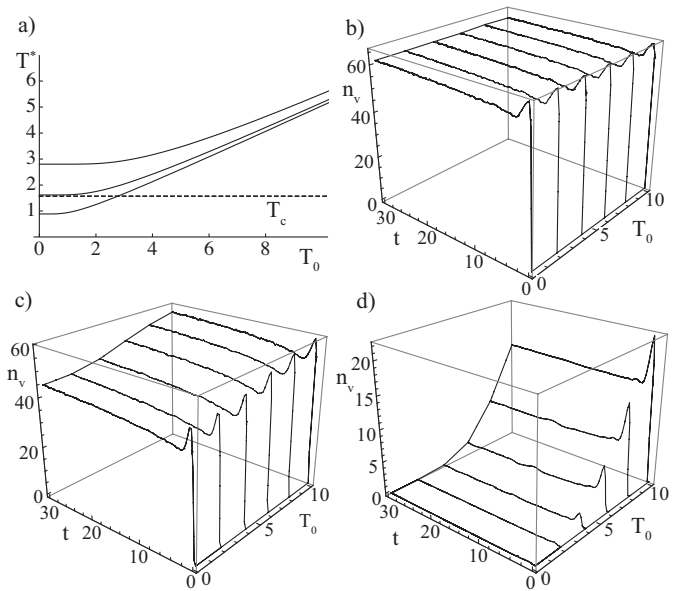}
\caption{\label{quenchsim}
a) $T^*$, as given in Eq. \ref{Tstarapp},
for $\kappa=1,3,10$, from top to bottom, and for
$V=20$. The line $T_c=\pi/2$ was added to indicate the critical
 temperature.
b) -- d) Simulations
for these values of $\kappa$ and $V$.
 We plot the number of vortices $n_v$ as a function of time $t$ 
 and initial temperature $T_0$. 
}
\end{figure}

The temperature that emerges inside the light cone can be estimated by considering the quadratures of $\phi$ at long times:
\bea
\lav\phi_k^2({t} \rightarrow \infty)\rav
& = & \frac{r_{k,0}}{4\omega_{k,0}}
 + \frac{r_{k,0} \omega_{k,0}}{4\omega_k^2}.
\eea
We find that the whole Wigner function in the non-interacting evolution remains Gaussian. It means that for each mode the Wigner function is equivalent to that of a harmonic oscillator at finite 'temperature'
$T^*_k$, which is in general mode-dependent:
\bea
 \frac{r_k^*}{\omega_k} & = & \frac{r_{k,0}}{2\omega_{k,0}}
 + \frac{r_{k,0} \omega_{k,0}}{2\omega_k^2}
\eea
where $r_k^* = 1/\tanh(\omega_k/2T^*_k)$.  Solving for $T_k^*$ gives
\bea\label{Tstarfull}
T^*_k & = & \frac{\omega_k}{2 \tanh^{-1}\Big(\frac{2\omega_k\omega_{k,0}}{
\omega_k^2 + \omega_{k,0}^2} \tanh(\omega_{k,0}/2T)\Big)}.
\eea
 For large $V\beta^2$, which corresponds to initially strong coupling between two superfluids, this simplifies to a single value, independent of  $k$:
\bea\label{Tstarapp}
T^* & = & \frac{\sqrt{V \beta^2}}{4 \tanh(\sqrt{V\beta^2}/2T)}.
\eea
For small initial temperatures $T$ we have $T^* \approx \sqrt{V
  \beta^2}/4$ ($T^\star=2J_\perp/J$ in terms of the original Hubbard
parameters), that is, the temperature is fully determined by the
initial coupling energy. The coupling energy between the two layers is
transferred into the in-plane kinetic energy.  We remind again that
this result is valid as long as $J_{\perp}\lesssim Un$, otherwise the
dependence of $T^\star$ on $J_{\perp}$ saturates and for the infinite
coupling limit we have $T^\star\sim Un/J$. For large $T$ we have $T^*
\approx T/2$.  This result is a reflection of the doubling of the
degrees of freedom when two layers are uncoupled.
In Fig. \ref{quenchsim} a) we show dependence $T^\star(T)$ evaluated
according to Eq.~(\ref{Tstarapp}), for $V=20$, and for $\kappa=1,3,10$
corresponding to lowering $J_\perp$.  For $\kappa=1$, $T^*$ is always
above the critical temperature $T_c =\pi/2$, for $\kappa=10$, it
crosses it.  We therefore expect to see very little vortex formation
for small temperatures for $\kappa=10$, and many vortices for all
temperatures for $\kappa=1$. The intermediate value $\kappa=3$
approximately describes the transition between these limits.
To make this point more clear in Fig. \ref{quenchsim} b)--d) we plot
full nonlinear TWA simulation for each of the three cases. We run the
quench for different temperatures $T$, and plot the number of vortices
$n_v$ in the system as a function of time. This number is obtained by
counting the vortices (indicated by red plaquettes in Fig.~\ref{run}),
and then by averaging over many runs. We find that for $\kappa=1$ the
number of vortices is virtually unchanged implying that the dynamics
is completely dominated by quantum fluctuations, whereas for
$\kappa=10$ this number drops to zero when the temperature is
lowered. These results are consistent with the emergent temperature
$T^*$ obtained within the linearized approach.

\section{Light-cone dynamics in the transverse Ising model}\label{tIsing}
In this section we demonstrate that light cone dynamics is not
just characteristic for the system we are interested in, which is
characterized by low energy bosonic wave excitations. The same
mechanism of reaching a steady state is much more general and is
likely related to the existence of the maximum group velocity in
Schr\"odinger systems as was proven by Lieb and
Robinson~\cite{Lieb-Robinson}. In this section we demonstrate the
presence of the light-cone dynamics in another solvable model, the
transverse Ising chain \cite{cardy, sachdev} described by the
Hamiltonian
\bea
H_I & = & -J_I \sum_i (\sigma^x_i \sigma^x_{i+1} + g \sigma^z_i),
\eea
where $J_I$ is an overall energy scale, $g$ describes the strength of the  transverse field, and $\sigma^{x,z}$ are the Pauli matrices.  We follow the calculational procedure in Ref.~[\onlinecite{sachdev}]. First, we use a Jordan-Wigner transformation
\bea
\sigma^z_i & = & 1 - 2 n_i\\
\sigma_i^x & = & \prod_{j<i} (1 - 2 n_j) (c_i + c^\dagger_i),
\eea
where  $c_i$ are Fermi operators, and $n_i = c_i^\dagger c_i$.  This transformation leads to a fermionic representation of the Hamiltonian, that can be further diagonalized  using the Bogoliubov transformation
\bea
\gamma_{k,g} & = & u_{k,g} c_k - i v_{k,g} c^\dagger_{-k}
\eea
where $c_k$ is the Fourier transform of $c_i$,  $u_{k,g}$ and $v_{k,g}$ are given by $\cos(\theta_{k,g}/2)$ and
 $\sin(\theta_{k,g}/2)$, where $\theta_{k,g} = \arctan(\sin k /(g-\cos k))$.  The resulting dispersion is
\bea\label{Isingdisp}
\epsilon_{k,g} & = &  2 J_I \sqrt{g^2 - 2 g \cos k + 1}.
\eea
\begin{figure}
\includegraphics[width=8cm]{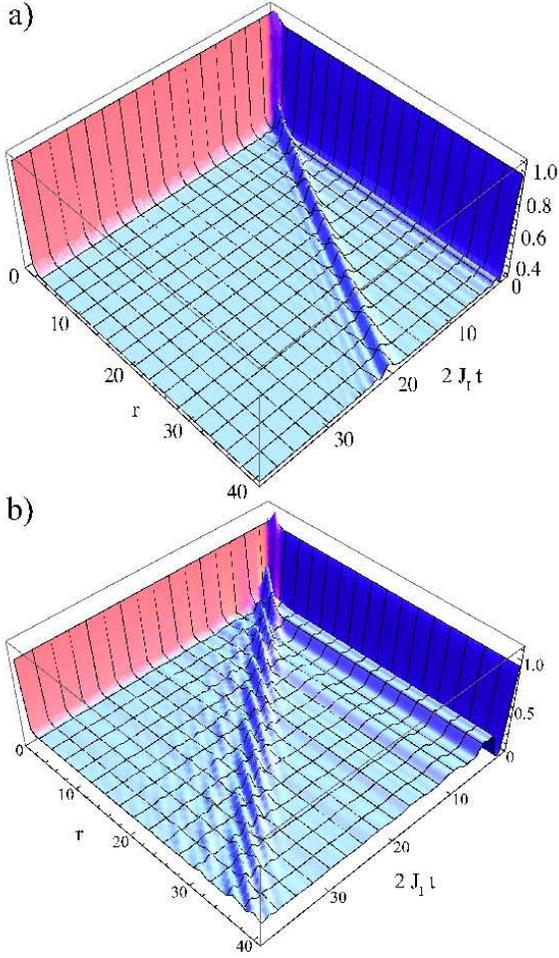}
\caption{The correlation function $\langle\sigma_0^z(t)
  \sigma_r^z(t)\rangle$ for a quench from $g=3$ to $g'=1$, in (a), and
  from $g=3$ to $g'=0.5$, in (b), as a function of time $t$,
  specifically of $2 J t$, and the spatial distance
  $r$.}\label{corr305}
\end{figure}
We consider a time dependent $g(t)$. For $t<0$ we have $g(t)=g$, and
we assume the system to be in equilibrium. We then assume that for
$t>0$, $g(t)$ jumps to the value $g'$.  The equal-time correlation
function of $\sigma^z_i$ can be calculated exactly by expressing it in
terms of the operator $n_i$, i.e.  $\lav \sigma^z_i(t)
\sigma^z_j(t)\rav = 1 - 4\lav n_i(t)\rav + 4 \lav n_i(t) n_j(t) \rav$.
It can be shown that the average density fermionic density
(corresponding to the $z$-component of the magnetization) is given by
\bea
\lav n_i(t)\rav & = & \lav n_i(0) \rav + \frac{1}{M} \sum_k F_{g,g'}(k,t)
\eea
with
\be
F_{g,g'}(k,t) = \frac{(\cos(2 \epsilon_{k,g'}t/\hbar)/2 - 1/2)(g'-g)\sin^2k}
 {\sqrt{g^2 - 2 g \cos k + 1} (g'^2 - 2 g'\cos k +1)}
\ee
and
\be
\lav n_i(0) \rav = \frac{1}{2} - \frac{1}{2M}\sum_k
\frac{g-\cos k}{\sqrt{g^2 - 2 g \cos k +1}}.
\ee
In turn the density-density correlation function $\lav n_i(t) n_j(t)\rav$ reads
\begin{widetext}
\bea
&&\lav n_i(t) n_j(t)\rav =  \frac{1}{M^2} \sum_{k_1, k_2}
\Big(\exp(-i(k_1 - k_2)(r_i - r_j))
 \Big((v^2_{k_1, g} + F_{g,g'}(k_1,t))(u^2_{k_2, g} - F_{g,g'}(k_2,t))\nonumber\\
&&~~~ +  (u_{k_1,g} v_{k_1,g} +G_{g,g'}(k_1, t))(u_{k_2, g} v_{k_2, g} +
 G^*_{g,g'}(k_2,t))
\Big)
+ (v^2_{k_1, g} + F_{g,g'}(k_1,t))(v^2_{k_2, g} + F_{g,g'}(k_2,t))\Big),
\eea
where
\be
 G_{g,g'}(k, t)=\Big(i \sin(2\epsilon_{k,g}t/\hbar)
 +\frac{1}{2}
  \frac{(\cos(2 \epsilon_{k,g'}t/\hbar) - 1)(g'-\cos k)}{\sqrt{g'^2 - 2 g'\cos k + 1}}\Big)
\Big(\frac{(g'-g)\sin k}{\sqrt{(g^2 - 2 g \cos k + 1)(g'^2 - 2 g'\cos k +1)}}\Big).
\ee
\end{widetext}
Using these expressions we can easily analyze the quench dynamics. In
Fig.~\ref{corr305} we show two examples showing spin-spin
(density-density) correlation function after a quench.  The first
example corresponds to the ramp from $g=3$ to $g'=1$, i.e. a quench to
the quantum critical point. At this point the dispersion
(\ref{Isingdisp}) becomes gapless and linear at small energies. Then
the light cone dynamics is anticipated because there is a well defined
``speed of light'' characterizing the propagation of excitations,
which is equal to $2J$. Indeed Fig.~\ref{corr305}a) shows clear
signature of such dynamics.  There is a clearly visible ``light
cone'', which separates into an instantaneous part (connecting not
causally connected points) that is independent of the distance, and a
spatially dependent (causal) part, that expands in the form of a wave
front.  In Fig. \ref{corr305} b) we use $g'=0.5$, where at low
energies the spectrum of excitations is gapped. Although the
dispersion in this model is linear (relativistic) only at sufficiently
high energies above the gap we still see a clear light-cone structure.
The expansion velocity of the ``light cone'' in this case is
consistent with twice the maximum of the group velocity $v_{gr}(k) =
d\epsilon/dk$ given by
\bea
v_{gr,max} & = &\left\{\begin{array}{ll}
2 J g & \mbox{for $|g|<1$}\\
 2 J & \mbox{for $|g|\geq 1$.}
\end{array}
\right.
\eea
So for $g'=1$ we find $2 v_{gr,max}=4 J$, and for $g'=0.5$ we find $2
v_{gr,max} = 2 J$. These are indeed the expansion velocities that we
see in Fig. \ref{corr305}.

\section{Dynamical vortex unbinding}\label{vunbind}
In this section we address the important question of how the
supercritical state relaxes to the ground state, i.e. the second stage
of the dynamics. As we mentioned in the introduction the anticipated
mechanism for this relaxation is vortex unbinding. This process is
intrinsically nonlinear and requires a more sophisticated treatment than
that of the noninteracting ``light cone'' dynamics. In this work we
use two complimentary approaches. In Sec.~\ref{vunbindnum} we use a
numerical implementation of the TWA to simulate the dynamics in the
system. In Sec.~\ref{vunbindRG} we generalize a renormalization group
approach to analytically describe the process of relaxation in real
time.

\subsection{Numerical approach}\label{vunbindnum}

\begin{figure}
\includegraphics[width=8.0cm]{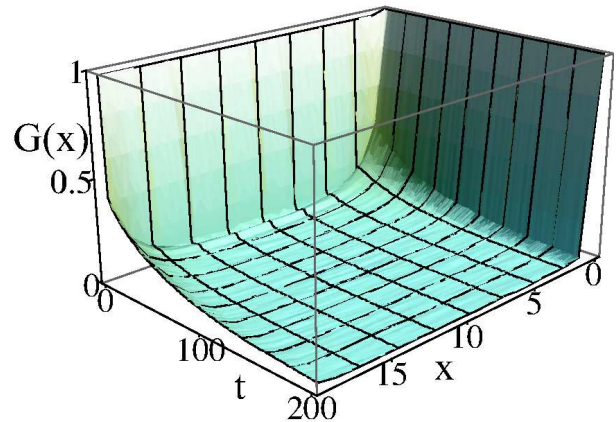}
\caption{\label{GFlong} Long-time behavior of the correlation function
  for $T=1$, $\kappa = 8$, and $V=80$.  The correlation function first
  develops algebraic scaling, so the system forms a metastable
  quasi-superfluid state.  On longer time scales the correlation function
  shows exponential decay.  The coherence is lost due to dynamical
  vortex unbinding.  }
\end{figure}

Within TWA we need to solve the full nonlinear equations of
motion~(\ref{eom}) subject to the initial conditions distributed
according to the Wigner function~(\ref{wig}). Then the equal-time
correlation functions or other observables are found by averaging the
Weyl symbol of the corresponding observable computed at time $t$ over
the fluctuating initial conditions. Note that since we are interested
only in phase-phase correlation function the corresponding Weyl symbol
is obtained by simply substituting the Heisenberg quantum operator
corresponding to the phase with the classical phase~\cite{ap_long}. In
Fig. \ref{GFlong} we show the result of such simulations. We can
observe how the metastable superfluid state relaxes to the disordered
state. For that, we show the correlation functions of the system on a
much longer time scale than in Fig.~\ref{GF}. The exponent of the
algebraic scaling gradually decreases. Eventually the correlation
function is more accurately approximated by an exponential fitting
function, signalling that the thermal Bose gas phase has been
reached. Because this is the phase of deconfined vortices, and because
the intermediate superfluid phase is well described by a phonon-only
description, we conclude that the dynamical transition that we observe
is due to vortex unbinding.  The example of a single run shown in
Fig. \ref{run} is consistent with this picture: Defects are created
soon after the quench, but they only gradually separate on a much
longer times scale. It is this process that we refer to as the
 reverse Kibble-Zurek mechanism.

To better characterize the process of vortex unbinding further we fit
the correlation function $G(x,t)$ to either algebraic or exponential
fitting functions. Such choice is motivated by the two possible
regimes of the equilibrium system and is supported by the analytic
renormalization group results presented in the next section. The
algebraic fitting function we use is $c(L/\pi|\sin(\pi
x/L)|)^{-\tau/4}$ and the exponential function is $c \exp(-|\sin(\pi
x/L)|/x_0)$.  Note that in the fitting functions we use the conformal
distance $L/\pi|\sin(\pi x/L)|)$, which is more appropriate in finite
systems with periodic boundary conditions~(see
e.g. Ref.~[\onlinecite{sachdev}]). In equilibrium the algebraic
exponent $\tau$ would be the relative temperature $T/T_c$. Any value
above $1$ is therefore supercritical.  The parameter $x_0$ defines the
length scale of the exponential decay.  The parameter $c$ in both
functions gives an overall scale.

\begin{figure}
\includegraphics[width=6.6cm]{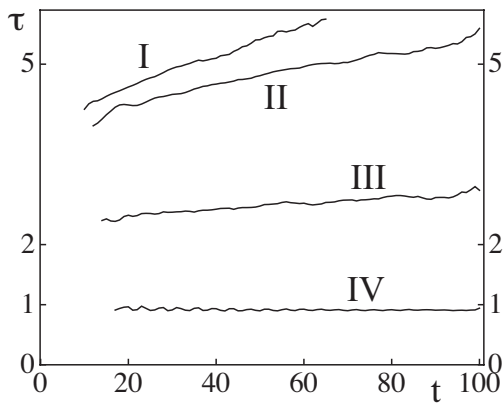}
\caption{\label{fit} Time dependence of the exponent $\tau$ extracted
  from fitting the long-time correlation function $G(x,t)$, for
  different initial couplings. In all four examples we use $T=1$ and
  $\kappa = 8$.  The initial couplings $V$ are chosen as $V=80, 70,
  50$, and $20$, corresponding to curves I to IV.  The curve I
  corresponds to the example shown in Fig. \ref{GFlong}.  In this case
  the correlation function can be well fitted with an algebraic
  function for up to $t \approx 60$, after that $G(x,t)$ is better
  fitted by an exponential function with a decay length of the order
  of the lattice constant. For the other cases, $G(x,t)$ is well
  fitted with an algebraic function throughout the whole time
  interval.}
\end{figure}

Using these fitting functions we analyze four different situations
corresponding to the same initial temperature $T=1$ and the same
parameter $\kappa=8$, but with different initial couplings $V$ between the
planes. The first (I) case corresponding to $V=80$ is identical to
the one plotted in Fig. \ref{GFlong}. The other three curves
correspond to $V=70, 50, 20$ (II--IV).  In Fig.~\ref{fit} we show the
exponent $\tau$ extracted from the fit as a function of time for these
situations. In all of them at short times $G(x,t)$ develops algebraic
scaling when the light-cone dynamics reaches the system boundaries. For
the cases I--III the emerging scaling exponent $\tau$ is well above
the critical exponent. After that, the exponent gradually increases on
much longer time scales. During this process, the decay of the
correlation function is still fitted well with the algebraic
function. Eventually the algebraic scalings breaks down and
$G(x,\tau)$ develops exponential scaling, indicating vortex unbinding.
This regime of exponential scaling is reached for $V=80$ (I) within
the time interval shown in Fig. \ref{fit}.  For $V=70$ (II) and $V=50$
(III) the time scale of the vortex unbinding is longer then the time
interval shown.  For $V=20$ (IV) the system equilibrates to the
superfluid state. Because in this case the exponent $\tau$ is less
than one, vortices never unbind and the algebraic scaling persists at
all times.  We conclude from these examples that there can be a
sizeable range of initial values of $V$ which generates the scenario
of a supercritical superfluid, and of dynamically suppressed vortex
unbinding. Furthermore, the algebraic scaling exponents that can occur
in the metastable state are well above criticality, and should be
easily distinguishable from subcritical values. These supercritical
exponents can be detected using interference experiments along the
lines of Refs.~\onlinecite{zoran, interference}.

\subsection{Renormalization group approach}\label{vunbindRG}
In this section we develop the renormalization group (RG) approach to
dynamical vortex unbinding. We find that the dynamical evolution of
the system can be related to the RG flow of the {\it equilibrium}
system. The idea of RG in real time is quite similar in spirit to the
RG in imaginary time. Namely our goal is to eliminate high energy,
high momentum degrees of freedom. In equilibrium, this is done by the
means of usual perturbation theory (or Gaussian integration), which is
justified because of the large energy gap separating high energy
states from the low-energy degrees of freedom we are interested in. In
real time the idea of renormalization is quite similar. High energy
(momentum) phonons are not very sensitive to slow processes leading to
vortex formations. Thus these phonons can be well treated within the
linearized approach. However, due to nonlinearities such phonons
slightly renormalize the parameters governing dynamics of low energy
degrees of freedom. This renormalization is precisely what we are
interested in. Note that technically in the RG procedure we perform
averaging of the equations of motion over short times. Then odd powers
of highly oscillating fields average to zero while averaging of the
even powers gives some constant contribution. This contribution is
precisely what renormalizes coupling constants governing the low
temperature dynamics.

We point out that typical RG flow diagrams contain mostly
non-equilibrium points, in fact, all except for the fixed points. As
we have seen in the previous sections, one can associate an effective
temperature to the metastable state that emerges after the dephasing
of the phonon modes. In turn with this effective parameter we can
associate a location of the transient state in the RG flow of the
equilibrium system. This effective temperature can then either
gradually increase, until the system starts to show exponential
scaling, or the system can always remain superfluid, if the algebraic
scaling is subcritical and the effective temperature always remains
below $T_{KT}$. This behavior resembles the equilibrium RG flow of a
Kosterlitz-Thouless transition (which now occurs in real, not
imaginary, time), on which we elaborate in this section.

Instead of directly analyzing the rotor model to describe the
Kosterlitz-Thouless physics and vortex unbinding, we will work
with the dual $Z_1$ clock model (or equivalently 2D sine-Gordon
model), described by the action
\bea\label{Ssg}
S & = & \int d^2 r \Big(\frac{\lambda}{2}(\partial_x \theta)^2
 - \frac{g}{a^2} \cos\theta\Big). 
\eea
For the details of the duality transformation see Ref.~[\onlinecite{wen}]. The parameters of this model can be related to those of the XY model by
\bea
\lambda & = &
 \frac{1}{8\pi} \frac{T}{T_{KT}}
=  \frac{1}{4\pi^2} \frac{T}{J_{KT}}\label{lambda}\\
\frac{g}{2} & = & \exp(- S_c),
\eea
where $E_c=S_cT$ is the vortex core energy, and $\lambda$ is a measure
of the relative temperature.  We note that the action in Eq. \ref{Ssg}
has a high-momentum cut-off $\Lambda$, which is the inverse of the
short-range cut-off $a$, i.e. we set $\Lambda a =1$.  To describe the
dynamics of this model we use the effective 2D sine Gordon
Hamiltonian:
\bea
H/T & = & \int d^2 r \Big( \frac{\mu}{2} p^2 - \frac{\lambda}{2}(\partial_x \theta)^2
 + \frac{g}{a^2} \cos\theta\Big).
 \label{hsg}
\eea
Here the parameter $\mu$ is chosen, so that the dispersion of the
linearized XY model is recovered:
\bea
\mu & = & \frac{\omega^2_k}{\lambda k^2 T^2},
\eea
which we also write as $\omega_k = v |k|$, where the
 velocity $v$ is given by $v=\sqrt{\mu\lambda T^2}$.
The nonlinear term
$\cos\theta$ in Eq.~(\ref{hsg}) describes the vortex field. If this
term is important (large $g$) then the field $\theta$ localizes
corresponding to a highly disordered phase of the dual field $\phi$,
i.e. to the normal state. Conversely small $g$ corresponds to the
superfluid algebraic regime. The starting point of our RG analysis
will be supercritical superfluid state which emerges after short time
light cone dynamics. Because the Kosterlitz-Thouless transition is
classical in nature occuring at high temperatures the quantum
fluctuations are no longer expected to be important and instead of
Wigner function as the new initial condition we can use its classical
Boltzmann's limit. The initial state for the vortex dynamics,
described by the effective temperature $T$, is thus fully
characterized by the quadratures of the spectrum:
\bea
\langle\theta^*_\bk\theta_\bk\rangle & = & \frac{1}{\lambda k^2},\\
\langle p_\bk^* p_\bk\rangle & = & \frac{1}{\mu} =
\lambda T^2 \frac{k^2}{\omega_k^2}.
\eea
The equations of motion corresponding to the Hamiltonian~(\ref{hsg}) are given by
\bea
\frac{d}{d t} p & = & \lambda T \partial_x^2 \theta
 + \frac{g T}{a^2} \sin\theta,\label{EOM1}\\
\frac{d}{d t} \theta & = & \mu T p.\label{EOM2}
\eea
We now apply the following renormalization procedure to these
equations.  We rescale the spatial and temporal variables as $\br
\rightarrow \br(1+ d\Lambda/\Lambda)$ and $t \rightarrow t(1+
d\Lambda/\Lambda)$, and the $p$-field as $p \rightarrow p(1-
d\Lambda/\Lambda)$.  This implies that the momentum cut-off $\Lambda$
is rescaled as $\Lambda \rightarrow \Lambda' \equiv\Lambda(1-
d\Lambda/\Lambda)$, so the momentum degrees of freedom between
$\Lambda'$ and $\Lambda$ are removed.  Without the non-linear term in
Eq. \ref{EOM1} these rescalings leave the equations of motion
invariant.  The linear dynamical evolution can therefore be considered
to be the non-interacting fixed point of the RG.  We now ask the
question, how this dynamical evolution is affected by the non-linear
term.  Specifically we want to determine how the equations of motion
behave at long times and distances.  For this, we go beyond
 the bare rescaling and correct for the integrated-out
 degrees of freedom up to second order in $g$.
 The resulting flow equations are of the well-known BKT form:
\bea
\frac{d g}{d l } & = & \Big(2 - \frac{1}{4\pi\lambda}\Big)g\label{RG1}\\
\frac{d\lambda}{dl} & = & \alpha \frac{g^2}{\lambda}.\label{RG2}
\eea
where $l = \ln \Lambda$, and $\alpha$ is a non-universal
prefactor. The RG step generated the equations of motion at time $t'$
and distance $r'$ from the equations at time $t$ and distance $r$,
with renormalized coefficients, according to Eqs. \ref{RG1} and
\ref{RG2}.  Therefore the time dependence of the coefficients can be
read off the solution of the RG flow, by realizing that: $dt/dl = t$
or $t = t_0 e^l$. In Fig. \ref{RGsketch} we show a schematic
representation of our RG process.  In the Appendix we discuss the
derivation of the flow equations and give their more complete form.
\begin{figure}
\includegraphics[width=9.0cm]{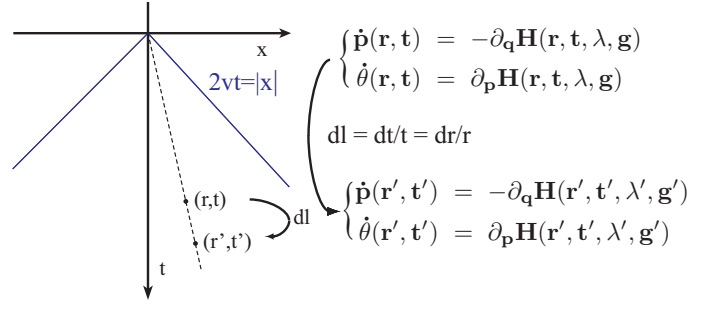}
\caption{\label{RGsketch}
 Schematic representation of a renormalization step in the real-time
 RG approach. In each step we renormalize simultaneously the space and time
 variables. This 'moves' the
 equations of motion from $(\br, t)$ to $(\br', t')$.
 We correct for the integrated-out degrees of freedom to second order
 in $g$, which renormalizes the  parameters $g$ and $\lambda$
 according to Eqs. \ref{RG1} and \ref{RG2}.
}
\end{figure}

One conclusion from Eqs. \ref{RG1} and \ref{RG2} is that the critical
exponent of the dynamical process is equal to the one of the
equilibrium system.  We see from Eq. \ref{RG1} that the critical value
of $\lambda$ is $\lambda_c = 1/8\pi$, which corresponds to $T=T_{KT}$
as can be seen from Eq. \ref{lambda}. Another important observation is
that the RG equations~(\ref{RG1}) and (\ref{RG2}) predict a
non-equilibrium analogue of the BKT transition, where depending on the
initial fluctuations in the system, the vortex-antivortex pairs can
either unbind in the long time limit or remain bounded. This
transition, as in the equilibrium case, is characterized by 
exponentially divergent time and length scales. Physically these
divergencies correspond to a very slow process of equilibration of
vortices near the nonequilibrium phase transition.

We can also use the RG flow to determine
 the time scale of vortex unbinding by using
\bea
g(t^*) & \sim & 1.
\eea
When $T$ is well above $T_{KT}$, the time scale can be determined
 from Eq. \ref{RG1},
\bea
t^* & \sim & \exp\Big(\frac{E_c/2}{T-T_{KT}}\Big)
\eea
where $E_c = S_c T$. Away from the transition, the
 time scale of vortex unbinding is therefore exponentially increased,
 because of the energy cost given by the vortex core energy.
 Very close to the transition  $t^*$ scales as:
\bea
t^* & \sim & \exp(\exp(-S_c/2)/\sqrt{1 - T_{KT}/T}).
\eea
 The time scale is renormalized because of the critical
 scaling in the vicinity of the transition.

\section{Conclusions}
In conclusion, we have studied the dynamics of the relative phase of a
bilayer of superfluids in 2D, after the hopping between them has been
turned off rapidly.  We find that on short time scales the dynamics of
the correlation function shows a 'light-cone'-like behavior. Depending
on the parameters of the system, the light cone dynamics can result in
a phase that shows supercritical algebraic scaling, and can therefore
be thought of as a superheated superfluid. On long time scales the
system relaxes to a disordered state via vortex unbinding, which
constitutes a reverse-Kibble-Zurek mechanism.  The properties of the
dynamical process can be understood with a renormalization group
approach. We find that the dynamical evolution of the system resembles
the RG flow of the equilibrium system. In particular, using the RG
equations we found two possible scenarios of the system reaching the
steady state: (i) if initial quantum and thermal fluctuations are weak
the vortices are irrelevant and long time long distance behavior is
governed by the algebraic fixed point. The only role of vortices is
then renormalization of the superfluid stiffness and the sound
velocity. (ii) If the initial fluctuations are strong then the
vortices become relevant and proliferate resulting in a normal
(non-superfluid) steady state. In this case RG gives the time scale of
vortex unbinding, which exponentially diverges as the system
approaches the non-equilibrium phase transition. The behavior of the
relative of phase of two superfluids can be accurately studied by
interference experiments of ultra-cold atom systems, and 
therefore our predictions are of direct relevance to experiment.

 \acknowledgments
We thank A. Castro Neto for useful discussions. Work of A.P. was supported by NSF DMR-0907039, AFOSR, and Sloan Foundation.
 L.M. acknowledges
 support from NRC/NIST, NSF Physics Frontier Grant PHY-0822671 and
 Boston University visitor's program.

\appendix\section{}
In this Appendix we derive the RG Eqs. \ref{RG1} and \ref{RG2},
 which can also be written as a second order differential
 equation for $\theta$
\bea
{1\over\mu}\frac{d^2}{d t^2} \theta & = & \lambda \triangle \theta
 + \frac{g}{a^2} \sin\theta.
 \label{eq:theta_1}
\eea
 To simplify the derivation, here and throughout the Appendix, we
 formally
 change notations $\lambda T\to
\lambda$, $\mu T\to \mu$, and $gT\to g$. The idea of momentum
shell RG is that we treat high momentum components of $\theta$ and $p$ (or equivalently $\dot\theta$)
perturbatively, while not making any approximations about the low momentum components. Our goal is to find renormalization of the equations of motion governing the low momentum components. So we split
\be
\theta({\bf r}, t)=\theta^<({\bf r},t)+\theta^>({\bf r},t),
\ee
where the Fourier expansion of $\theta^>({\bf r},t)$ only contains
momenta in the shell $\Lambda'\equiv\Lambda-\delta \Lambda<|k|<\Lambda$ and
$\theta^<({\bf r},t)$ contains all other Fourier components:
\bea
 \theta^<(\br)& = & \frac{1}{\sqrt{V}}\sum_{k<\Lambda'} \exp(i\bk \br) \theta_\bk\\
 \theta^>(\br) & = &  \frac{1}{\sqrt{V}}\sum_{\Lambda'<k<\Lambda} \exp(i\bk \br) \theta_\bk.
\eea
We will
treat $\theta^>$ (and correspondingly $p^>$) perturbartively in $g$
since the nonlinear term should only weakly couple to the high
frequency field.
 We expand the high-momentum field as
\be \theta^>({\bf k},t)=\theta_0^>({\bf
  k},t)+\theta_1^>({\bf k},t).
\ee
Here $\theta_0^>(\bk, t)$
 is the solution of the equations
 of motion, with $g$ set to zero:
 \be \theta_0^>({\bf
  k},t)={\mu\over\omega_\Lambda}p_{0,{\bf k}}^>\sin(\omega_\Lambda
t)+\theta_{0,{\bf k}}^>\cos(\omega_\Lambda t), \ee where
$\omega_k= v|k|$, and the
 velocity $v$ is $v = \sqrt{\lambda\mu}$. In the next leading order we have \be
\theta_1^>({\bf k},t)={g\omega_\Lambda\over \lambda}\int_0^t
d\tau F_1({\bf k},\tau)\sin(\omega_\Lambda(t-\tau)),
\label{theta_1}
\ee
where
\be
F_1({\bf k},\tau)=\int d^2 r \exp[-i{\bf k r}]\sin(\theta_0^<({\bf r},\tau)),
\label{f_tau}
\ee
 and we used $\Lambda a = 1$.
Note that in the last equation in the argument of the sinus we changed $\theta_0$ to $\theta_0^<$ because the contribution from $\theta_0^>$ is smaller by the factor $\delta\Lambda/\Lambda$. So we see that in the leading order in $g$ the high momentum component of $\theta$ oscillates with time at very high frequency $\omega_\Lambda$. In the next order in $g$ the high momentum component also acquires a low frequency component (as we will discuss below).

Next we consider the equation of motion (\ref{eq:theta_1}) expanding it up to the second order in $\theta^>$:
\bea
{1\over\mu }{d^2\over dt^2}\theta({\bf r},t)&\approx&\lambda\Delta\theta({\bf r},t)+ \frac{g}{a^2}\cos(\theta^<({\bf r},t))\theta^>({\bf r},t)\nonumber\\
&+& \frac{g}{a^2}\sin\theta^<({\bf r},t)\left(1-{(\theta^>({\bf r},t))^2\over 2}\right).
\label{eq:p_<}
\eea
Because of the nonlinearity high-momentum modes couple to the low
momentum modes leading to the renormalization of the couplings
governing the dynamics of the latter. The idea of RG is to average
equations of motion for low-momentum (slow) components over the fast
oscillations. The averaging is trivially done in the last term of
Eq.~(\ref{eq:p_<}). There it is sufficient to use zeroth order in
$\theta^>$. Using that $\overline{\sin^2(\omega_\Lambda t)},
\overline{\cos^2(\omega_\Lambda t)}=1/2$ we find that averaging of the
last term simply renormalizes the coupling $g$: 
\be g\to
g\left(1-{\overline{ E_\Lambda}\over
    4\pi\lambda}{\delta\Lambda\over\Lambda}\right),
\label{eq:g}
\ee
where $\overline{E_k}$ is the average energy of the mode $k$ over the
period (we used the fact that $\lambda
k^2\overline{|\theta_k|^2}=\overline{E_k}$). 
 We note that in a Boltzmann ensemble, we would have $\overline{E_k} =1$,
 because the energies here are in units of the temperature $T$. 
 With this assumption we would recover the
 flow equation of the equilibrium case.

 Instead of this assumption, we proceed by noting that under RG
 transformations coupling constants slowly change in time. This
 implies that the adiabatic invariants per each mode are approximately
 conserved, as discussed in Ref. \onlinecite{LL1}. For an oscillator the
 adiabatic invariant is $I_k=E_k/\omega_k$. Thus we see that the
 energy of the mode is proportional to the frequency. Noting that at
 initial time $\overline{E_k}(t=0)=1$ (in non-rescaled units this
 would be $\overline{E_k}(t=0)=T_0$, where $T_0$ is the initial
 non-equilibrium temperature), one can rewrite Eq.~(\ref{eq:g}) as
 follows:
\be
g\left(1-{1\over 4\pi v_0}{1\over K}{\delta\Lambda\over\Lambda}\right),
\label{eq:g1}
\ee
where we introduced the analogue of the Luttinger-Liquid parameter
$K=\sqrt{\lambda/\mu}$. $v$ is the velocity which is now given by
$v=\sqrt{\lambda\mu}$ (note that in the original, not rescaled units,
$v=T\sqrt{\lambda\mu}$).

Next let us consider the second term in Eq.~(\ref{eq:p_<}). This term
is more subtle since if we use $\theta_0^>$ the average over fast
fluctuations will give zero. So we need to use the first correction
$\theta_1^>$, which would be a correction at second order in $g$.
We note that the linear term in Eq.~(\ref{eq:p_<}) can also  be expanded to second in $g$, generating similar contributions.  However, when written as Eq.~(\ref{eq:theta_1}),  such a term a cancelled by a corresponding term from expanding $d^2\theta/dt^2$. Alternatively we can view Eq.~(\ref{eq:p_<}) as written for the $\theta^<$ component, this automatically ensures that only the nonlinear term is responsible for the renormalization.

Let us look closer into the Eq.~(\ref{theta_1}). We are dealing with
the integral over the fast oscillating function of $\tau$:
$\sin(\omega_\Lambda(t-\tau))$ and the slow oscillating function
$F$. This integral can be evaluated by integrating by parts:
\bea
&&\int_0^t d\tau F_1(\tau)\sin(\omega_\Lambda(t-\tau))=F_1(\tau){\cos(\omega_\Lambda(t-\tau))\over \omega_\Lambda}\biggr|_0^t\nonumber\\
&&~~~~-{1\over\omega_\Lambda}\int_0^t d\tau {d F_1(\tau)\over d\tau} \cos(\omega_\Lambda(t-\tau)).
\eea
Note that the second integral contains a large denominator $1/\omega_\Lambda$. In the first term only the limit $\tau=t$ gives a nonoscillating contribution to the integral. We can continue the expansion in powers of $1/\omega_\Lambda$. Note that the next term proportional to $\dot F_1$ will contain only highly oscillatory part and can be neglected. So up to the third order in $1/\omega_\Lambda$ we find:
\be
\int_0^t d\tau F_1(\tau)\sin(\omega_\Lambda(t-\tau))\approx {F_1(t)\over \omega_\Lambda}-{1\over\omega_\Lambda^3}{d^2 F_1(t)\over dt^2}.
\label{int}
\ee
 Combining Eqs.~(\ref{theta_1}), (\ref{f_tau}), and (\ref{int}) we find
\bea
&&\theta_1^>({\bf k},t)\approx {g\over \lambda} \int d^2 r \exp[-i{\bf k r}]\sin(\theta^<({\bf r},t))\nonumber\\
&&-{g\over\lambda\omega_\lambda^2}\int d^2 r \exp[-i{\bf k r}]\cos(\theta^<({\bf r},t)) \ddot\theta^<({\bf r},t),
\eea
where we used $\Lambda a=1$ again.
Here we neglected by the term proportional to $(\dot\theta^<({\bf r},t))^2$, because it leads to a subdominant (in the RG sense) contribution. From this we find that
\begin{widetext}
\be
\theta_1^>({\bf r},t)\approx{g\over\lambda}\int\limits_{\rm shell} {d^2 k\over (2\pi)^2}\mathrm e^{i{\bf kr}} \int d^2x \mathrm e^{-i{\bf kx}}\sin(\theta^<({\bf x},t))-{g\over\lambda^2\Lambda^2\mu}\int\limits_{\rm shell} {d^2 k\over (2\pi)^2}\mathrm e^{i{\bf kr}} \int d^2x \mathrm e^{-i{\bf kx}}\cos(\theta^<({\bf x},t))\ddot\theta^<({\bf x},t).
\ee
 We now consider the term $(g/a^2)\cos(\theta^<(\br, t))\theta_1^>(\br, t)$:
\bea
\frac{g}{a^2}\cos(\theta^<(\br, t))\theta_1^>({\bf r},t)
& \approx&{g^2\over\lambda a^2}\int\limits_{\rm shell} {d^2 k\over (2\pi)^2}\mathrm e^{i{\bf kr}} \int d^2x \mathrm e^{-i{\bf kx}}\frac{1}{2}\sin(\theta^<({\bf x},t) -\theta^<({\bf r},t))\nonumber\\
&&-{g^2\over\lambda^2\mu}\int\limits_{\rm shell} {d^2 k\over (2\pi)^2}\mathrm e^{i{\bf kr}} \int d^2x \mathrm e^{-i{\bf kx}}\frac{1}{2}\ddot\theta^<({\bf x},t),
\eea
where we neglected terms such as $\sin(2\theta^<({\bf r},t))$.
Because we are integrating over high momentum shell this integral will
be suppressed unless ${\bf x}$ is close to ${\bf r}$. This suggests
change of variables ${\bf x}={\bf r}+{\bf \xi}$ and Taylor expanding
$\theta^<(\bx, t)$ in powers of $\bm{\xi}$, i.e.  $\theta^<(\bx, t)
\approx \theta^<({\bf r},t)+\bm{\xi\nabla}\theta^<({\bf r},t)+{1\over
  2}\xi_\alpha\xi_\beta {\partial \theta^<({\bf r},t)\over\partial
  r_\alpha\partial r_\beta}$. Then
\bea
\frac{g}{a^2}\cos(\theta^<(\br, t))\theta_1^>({\bf r},t)
& \approx&  \frac{C_2}{8\pi} \frac{g^2}{\lambda} \frac{\delta\Lambda}{\Lambda}
 \triangle\theta
 - \frac{C_1}{4\pi} \frac{g^2}{\lambda^2\mu}
 \frac{\delta\Lambda}{\Lambda} \ddot\theta,
\eea
\end{widetext}
where
\be
\frac{C_1}{\Lambda^2}=\int d^2\xi\, J_0(\Lambda\xi),\quad
 \frac{C_2}{\Lambda^4}=\int d^2\xi\,\xi^2 J_0(\Lambda\xi).
\ee
When we need to substitute these expressions back into
Eq.~(\ref{eq:p_<}), we find that the term containing $\triangle\theta$
renormalizes the coupling $\lambda$ as
\be \lambda\to\lambda +
\frac{C_2}{8\pi}{g^2\over \lambda}{\delta\Lambda\over\Lambda}.
\label{prefactor1}
\ee
In addition there is an extra term proportional to $\ddot\theta$
generated in Eq.~(\ref{eq:p_<}), which renormalizes $\mu$:
\begin{equation}
{1\over \mu}\to {1\over \mu}+\frac{C_1}{4\pi}{g^2\over \lambda^2\mu}{\delta\Lambda\over\Lambda}.
\label{prefactor2}
\end{equation}
%
%
%
Finally we restore the cutoff
by rescaling ${\bf k}\to {\bf k}(1-\delta\Lambda/\Lambda)$, ${\bf
  r}\to {\bf r}(1+\delta\Lambda/\Lambda)$, $t\to t
(1+\delta\Lambda/\Lambda)$, and $p\to
p(1-\delta\Lambda/\Lambda)$. This rescaling additionally renormalizes
the coupling $g$: $g\to g(1+2\delta\Lambda/\Lambda)$. Combining this result with Eqs.~(\ref{eq:g1}), (\ref{prefactor1}), (\ref{prefactor2}) we find the following renormalization group equations:
\bea
{dg\over dl}&=&g\left(2-{1\over 4\pi v_0}{1\over K}\right)\label{RG3}\\
{dK\over dl}&=&{1\over 16\pi K}{g^2\over v^2}(C_2+2C_1),\label{RG4}\\
{dv\over dl}&=&{g^2\over 16\pi v K^2}(C_2-2C_1)\label{RG5}.
\eea
where $l=\ln\Lambda$. We can read off from Eq. \ref{RG5}, that if 
 the system contains a fixed velocity, for example in relativistic systems,
  we need to have $C_2=2C_1$, to enforce that the velocity is
 invariant under the flow. 
 
Note that if the initial system is already close
to the critical point then the RG equations above simplify to \bea
{dg\over dl}&\approx&g\left(2-{1\over 4\pi\lambda}\right),\\
{d\lambda\over dl}&\approx& {C_2\over 8\pi} {g^2\over\lambda}, \eea
which are equivalent to Eqs.~(\ref{RG1}) and (\ref{RG2}). Note that a
more complete set of RG equations (\ref{RG3}) - (\ref{RG5}) has the
same universal predictions of the dynamical phase transitions and
exponential divergence of the time scales as the simplified equations
above. Also note that the real RG equations bear close analogy to the
flow equations in imaginary time characterizing the equilibrium
Kosterlitz-Thouless transition~\cite{giamarchi_book}. Thus the
non-equilibrium KT transition discussed here is characterized by
exponentially divergent length and time scales. Physically these long
scales characterize very slow process of vortex unbinding and
equilibration at long distances. Note that the RG
equations~(\ref{RG3}) - (\ref{RG5}) also implicitly take into account
renormalization of the temperature in the system. This comes from the
fact that creating vortex-antivortex pairs removes the energy from the
phonon degrees of freedom. We are going to investigate this issue in
more detail in a separate publication.

\def\etal{\textit{et al.}}

\end{document}